\documentstyle[aps,twocolumn,epsfig]{revtex}

\begin{document}
\title{\begin{flushright}
\footnotesize{CECS-PHY-00/12}\\ \footnotesize{USACH-FM-00/10}
\end{flushright}Degenerate Dynamical Systems}
\author{Joel Saavedra$^{1,2}$\thanks{${\tt joelium@cecs.cl}$}, Ricardo Troncoso$^{1}$\thanks{${\tt ratron@cecs.cl}$} and Jorge Zanelli$^{1}$\thanks{${\tt jz@cecs.cl}$}}
\address{$^{1}$Centro de Estudios Cient\'{\i }ficos (CECS), Casilla 1469, Valdivia, Chile.\\
$^{2}$Departamento de F\'{\i }sica, Universidad de Santiago de Chile, Casilla 307, Santiago 2,
Chile.}
\date{May 23, 2001}
\maketitle

\begin{abstract}
Dynamical systems whose symplectic structure degenerates, becoming
noninvertible at some points along the orbits are analyzed. It is shown that
for systems with a finite number of degrees of freedom, like in classical
mechanics, the degeneracy occurs on domain walls that divide phase space
into nonoverlapping regions each one describing a nondegenerate system,
causally disconnected from each other. These surfaces are characterized by
the sign of the Liouville's flux density on them, behaving as sources or
sinks of orbits. In this latter case, once the system reaches the domain
wall, it acquires a new gauge invariance and one degree of freedom is
dynamically frozen, while the remaining degrees of freedom evolve regularly
thereafter.
\end{abstract}

\section{Introduction}

A number of dynamical systems of physical interest possess field-dependent symplectic forms which
degenerate, becoming noninvertible for some particular configurations. Systems as diverse as
vortex interactions in fluids \cite{Aref}, and gravitation theories in dimensions $d>4$ containing
higher powers of curvature in the Lagrangian exhibit this feature (see e.g.
\cite{Teiltelboim-Zanelli}). Models of this kind naturally arise in different contexts of current
high energy physics, ranging from cosmology and brane worlds \cite{BW1,BW2} to strings an
M-theory \cite{CHSW,GV,TrZ}.

The problem is how to describe the evolution of the system near a degenerate
configuration and, if it could reach such state, how it would evolve
afterwards. The standard hypotheses in the treatment of dynamical systems,
however, exclude the possibility that the symplectic form may have
nonconstant rank throughout phase space, even in classical mechanics (see,
{\it e.g.}, \cite{Arnold,HT}).

As a first step towards understanding the general problem, here we analyze
degenerate dynamical systems in classical mechanics. We show that it is
possible to fully characterize the evolution of these systems.

It should be emphasized that this degeneracy is independent of Poincar\'{e}'s classification of
singularities. A Poincar\'{e} singularity occurs at critical points of the Hamiltonian, which are
generically isolated, whereas the symplectic form degenerates on surfaces which are generically
domain walls. This kind of surfaces cannot be understood as dense sets of Poincar\'{e}
singularities. Roughly speaking, a symplectic degeneracy is the counterpart of a Poincar\'{e}
singularity in that, in the latter the gradient of the Hamiltonian vanishes, whereas the former
can be interpreted as an infinite gradient.

The previous point can be made more explicit, by considering the simplest
example of a degenerate system, whose phase flow satisfies
\begin{equation}
\left(
\begin{array}{ll}
0 & x_{2} \\
-x_{2} & 0
\end{array}
\right) \left(
\begin{array}{l}
\dot{x}_{1} \\
\dot{x}_{2}
\end{array}
\right) =\left(
\begin{array}{l}
E_{1} \\
E_{2}
\end{array}
\right) \;,  \label{Simplest}
\end{equation}
with $E_{1}E_{2}\neq 0$, which degenerates at $x_{2}=0$. An equivalent
formulation in the $x_{2}\neq 0$ region is
\begin{equation}
\left(
\begin{array}{l}
\dot{x}_{1} \\
\dot{x}_{2}
\end{array}
\right) =\frac{1}{x_{2}}\left(
\begin{array}{l}
-E_{2} \\
E_{1}
\end{array}
\right) \;,\label{Simplest2}
\end{equation}
which can be viewed as a phase flow where the gradient of the Hamiltonian diverges as
$x_{2}\rightarrow 0$. The required symplectomorphism (canonical transformation) to obtain Eq.
(\ref{Simplest2}) from Eq. (\ref{Simplest}) is noninvertible throughout phase space, however.

\section{First-order Lagrangians and their Symplectic forms}

Let us consider a system whose action is a one-form $A$, integrated over a $%
(0+1)$-dimensional worldline embedded in a $(2n+1)$-dimensional spacetime of
signature $(-,+,...+)$,
\begin{equation}
S[z;1,2]=\int_{1}^{2}A_{\mu }\dot{z}^{\mu }d\tau \;,  \label{action}
\end{equation}
The field $A_{\mu }$ is a prescribed set of $2n+1$ functions of the
embedding coordinates $z^{\mu }$, which are the dynamical variables \cite
{HU+FJ}. This action is manifestly invariant under reparametrizations of the
worldline $\tau \rightarrow \tau ^{\prime }(\tau )$, and diffeomorphisms $%
z^{\mu }\rightarrow z^{\prime \mu }(z)$ \cite{Symplectomorphisms}.
Identifying the affine parameter with the timelike embedding coordinate $%
z^{0}:=t$, so that $z^{i}=z^{i}(t)$, the action reads
\begin{equation}
S[z;1,2]=\int_{t_{1}}^{t_{2}}[A_{i}\dot{z}^{i}+A_{0}]dt\;.  \label{action'}
\end{equation}
The equations of motion are given by
\begin{equation}
F_{ij}\dot{z}^{j}+E_{i}=0\;,  \label{E-L}
\end{equation}
where we have defined $E_{i}\equiv \partial _{i}A_{0}-\partial _{0}A_{i}$
and $F_{ij}\equiv \partial _{i}A_{j}-\partial _{j}A_{i}$. In the following,
we assume $A_{i}$ and $A_{0}$ to be time-independent.

These dynamical systems are naturally classified according to the rank $\rho
$ of the symplectic\footnote{%
The standard notion of symplectic form is usually assigned only to nonsingular closed two-forms.
However we extend the term ``symplectic form'' to the cases (B) and (C) below.} form $F_{ij}$.
Thus, three cases are distinguished: {\bf (A)} Regular Hamiltonian systems, for which the
symplectic form has constant maximal rank, $\rho (F_{ij})=2n$ throughout phase space $\Gamma $
\cite{HU+FJ}. {\bf (B)} Singular or constrained Hamiltonian systems, which have a constant
nonmaximal rank, $\rho (F_{ij})=2m<2n$ throughout $\Gamma $ \cite{HT}. And, {\bf (C)} Degenerate
systems, which have nonconstant rank $\rho (F_{ij})$ throughout $\Gamma $.

\section{Degenerate Systems}

We will focus our discussion in the degenerate case (C), which has been traditionally left aside
in the literature. We will assume that the zero-measure subset of $\Gamma $ given by
\begin{equation}
\Sigma =\{z\in \Gamma /F=0\}\;,
\end{equation}
where $F:=\det (F_{ij})$, is not dense. Thus, outside $\Sigma $, the
symplectic form $F_{ij}$ has a constant rank $2n$, and the dynamical
structure there is described through cases (A) above \footnote{%
The case in which $\rho (F_{ij})$ is less than maximal in the complement of $%
\Sigma $, is a combination of cases (B) and (C). It is straighforward to consider this additional
complication, but it does not add much to deserve an extensive discussion here.}.

Under these conditions, nothing prevents the system, starting from a generic
state for which $F\neq 0$, from reaching a point on $\Sigma $ after some
finite time. Having this scenario in mind, we address the following points:

$\bullet $ The description of the locus of $\Sigma $.

$\bullet ${\bf \ }Classification of the phase flow near $\Sigma $.

$\bullet ${\bf \ }Whether $\Sigma $ can be reached and, in that case, the
fate of the system thereafter.

\subsection{Degeneracy Surfaces $\Sigma $}

As is well known, a skew-symmetric $2n\times 2n$ matrix $F_{ij}(z)$ can be
brought into the block-diagonal form by an orthogonal transformation. Thus
the two-form ${\cal F}=\frac{1}{2}F_{ij}dz^{i}\wedge dz^{j}$ can be block
diagonalized in an open set, under a local $O(2n)$ coordinate transformation
$z^{i}\rightarrow x^{i}(z)$,
\begin{equation}
{\cal F}=\sum\limits_{r=1}^{n}f_{r}(z)dx^{2r-1}\wedge dx^{2r}\;.
\label{canonicalF}
\end{equation}

However,{\em \ in open sets containing points of the degeneracy surfaces,
the Darboux-like coordinates }$x^{i}${\em \ cannot be brought into the
standard canonical form, because at least one of the }$f_{r}${\em 's in }(%
\ref{canonicalF}){\em \ vanishes at }$\Sigma ${\em .} Hence, further
(finite) rescalings cannot normalize the $f_{r}$'s to $1$. As a consequence,
the set $\Sigma $ is the union of the $(2n-1)$-dimensional surfaces
\[
\Sigma _{r}=\{z\in \Gamma /f_{r}(z)=0\}\;,
\]
that is, $\Sigma =\cup _{r=1}^{n}\Sigma _{r}$.

Moreover, by virtue of the Bianchi identity ($d{\cal F}=0$), it can be shown
that $f_{r}(x)$ depends only on the pair of conjugate coordinates $%
(x^{2r-1},x^{2r})$. This means that the degeneracy surfaces are constant
along the remaining coordinates.

We assume that the $f_{r}$'s are smooth Morse functions on the corresponding
$(x^{2r-1}-x^{2r})$ planes, which ensures that they possess only simple
zeros except at isolated points; the cases where $f_{r}$ has zeros of higher
order can be thought of as the merging of simple zeros. Hence, the level
curves $f_{r}(x^{2r-1},x^{2r})=0$ divide the $(x^{2r-1}-x^{2r})$--plane into
nonoverlapping sets and therefore,\newline

{\bf Lemma 1:} The locus of the degeneracy surfaces $\Sigma $\ corresponds
to a collection of domain walls, splitting the phase space $\Gamma \,$into a
number of nonoverlapping regions.

\subsection{Characterization of the Phase Flow near $\Sigma $}

Generically, at a surface $\Sigma _{r}$ the rank $\rho (F_{ij})$ is lowered
by $2$, and at points where $k$ of these surfaces intersect, $\rho $ is
lowered by $2k$. In a sufficiently small neighborhood of the surface $\Sigma
_{r}$, the behavior of the system is dominated by the dynamical variables $%
x^{\alpha }=(x^{2r-1}-x^{2r})$, whose corresponding equations of motion can
be read from Eq. (\ref{E-L}) as
\begin{equation}
\epsilon _{\alpha \beta }f(x)\dot{x}^{\beta }=-E_{\alpha }\text{ },
\label{hamilton}
\end{equation}
where for simplicity, we have set $r=1$, so that $\alpha $ and $\beta =1,2$
and $f:=f_{1}$. Near a degeneracy surface $\Sigma _{r}$, the remaining
dynamical variables $z^{a}$, $(a=3,...,2n)$,\ behave like the phase space
coordinates of a regular system.

Here it is assumed that $E_{\alpha }$, remains finite and does not vanish on
$\Sigma _{1}$ (i.e., Poincar\'{e} singularities are assumed to be located
outside $\Sigma $), therefore, Eq. (\ref{hamilton}) implies that the
velocity becomes tangent to the $(x^{1}-x^{2})$ plane, because the
components $\dot{x}^{\alpha }$ become unbounded as the orbit approaches $%
\Sigma _{1}$, while the other components ($\dot{z}^{a}$) remain finite.

Due to the fact that $f$ has a simple zero at $\Sigma _{1}$, $\dot{x}%
^{\alpha }$ reverses its sign across the degeneracy surface. Consequently,
the phase flow evolves in opposite directions on each side of $\Sigma $.
Thus, in a local neighborhood of $\Sigma $, one of the following three
situations occur: (a) Orbits flow towards $\Sigma $ and end there, (b) the
orbits originate at the degeneracy surface and flow away from it, or (c) the
orbits run parallel to $\Sigma $, but in opposite directions on each side.

Hence, the surfaces act as sinks or sources for the orbits in cases (a) and
(b) respectively, which naturally suggests a classification of the local
nature of $\Sigma $ into $\Sigma ^{(-)}$, $\Sigma ^{(+)}$, and $\Sigma
^{(0)} $ for the cases (a), (b) and (c), respectively (see Fig. 1).

In all three cases there is no flux across the degeneracy surface, and
therefore,\newline

{\bf Lemma 2:} The regions on either side of $\Sigma $\ are causally
disconnected and dynamically independent from each other.\newline

An immediate consequence of this, is the violation of Liouville's theorem at
the surfaces of degeneracy. In fact, outside the degeneracy surfaces, the
Liouville current
\begin{equation}
j^{i}=\sqrt{F}\dot{z}^{i}\;,
\end{equation}
is divergence-free ($\partial _{i}j^{i}=0$) by virtue of the equations of
motion and the identity $\partial _{i}(\sqrt{F}F^{ij}E_{j})=0$, with $%
F^{ij}F_{jk}=\delta _{k}^{i}$. This means that Liouville's theorem holds
outside $\Sigma $, where the dynamical behavior is regular. Moreover, $j^{i}$
has a finite limit as the system approaches a degeneracy surface, whose only
nonvanishing components on each side of $\Sigma $ are
\begin{equation}
j^{\alpha }=|f|\dot{x}^{\alpha }=sgn(f)\epsilon ^{\alpha \beta }E_{\beta }\;.
\end{equation}

The local character of the degeneracy surfaces $\Sigma $, can be inferred
from the flux of $j^{i}$ across a pill box enclosing a portion of $\Sigma $.
The flux density $\Phi =j^{i}n_{i}$ across the lids of the pill box is given
by the projection of $j^{i}$ along the normal to the surface $n_{i}=\partial
_{i}F^{1/2}$, whose only non vanishing components are $n_{\alpha }=\partial
_{\alpha }|f|$, that is,
\begin{equation}
\Phi =-F^{1/2}F^{ij}E_{j}\partial _{i}F^{1/2}=\partial _{\alpha }f\epsilon
^{\alpha \beta }E_{\beta }\;.  \label{phi}
\end{equation}
Note that $\Phi $ is not only finite, but continuous on $\Sigma $. Therefore,%
\newline

{\bf Lemma 3:} The local character of the degeneracy surfaces is given by $%
\Sigma ^{(\eta )}$\ with $\eta =sgn(\Phi )$. Furthermore, in general,\ $%
\Sigma $\ is globally piecewise attractive ($\Sigma ^{(-)}$)\ or repulsive ($%
\Sigma ^{(+)}$), and is of type $\Sigma ^{(0)}$ at the intersections with
the surfaces $\Pi =\{z\in \Gamma /\Phi (z)=0\}$ (see Fig. 1.d).\newline
\begin{figure}[tbm]
\begin{center}
  \leavevmode
  \epsfxsize=3 in
\epsfbox{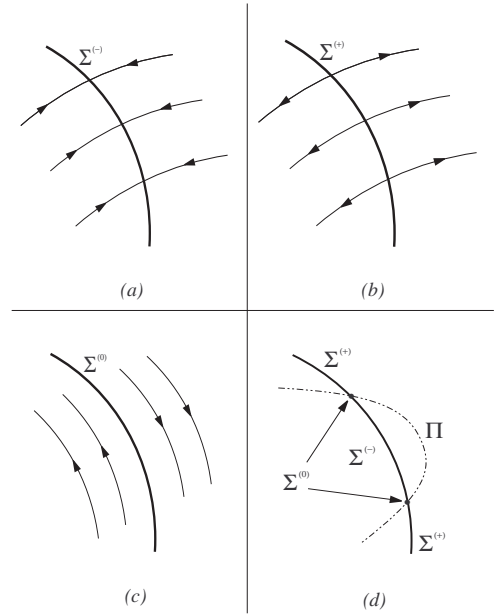}\label{DegSurfaces} \caption{(a), (b) and (c) show the qualitative local
flow in the neighborhood of $\Sigma ^{(+)}$, $\Sigma ^{(-)}$ and $\Sigma ^{(0)}$, respectively.
The global structure of degeneracy surfaces is shown in (d).}
\end{center}
\end{figure}
Hence, $\Sigma ^{(0)}$ generically corresponds to the boundaries between $%
\Sigma ^{(-)}$ and $\Sigma ^{(+)}$ (that is, $\Sigma ^{(0)}=\partial \Sigma
^{(-)}$) which is a subset of codimension $2$ in phase space.

In the particular case, when both surfaces $\Sigma $ and $\Pi $ coincide on
an open set, $\Sigma $ is globally of type $\Sigma ^{(0)}$. This occurs for
example, if $\left. E_{i}\right| _{\Sigma ^{(0)}}=\partial
_{i}(h(z^{i})F^{1/2})$, whose only nonvanishing components are of the form $%
E_{\alpha }=\tilde{h}(z^{a})\partial _{\alpha }f$ for some functions $h$ and
$\tilde{h}\neq 0$\ \cite{Dense}.

\subsection{Evolution towards $\Sigma ^{(-)}$}

The degeneracy surfaces $\Sigma ^{(+)}$and $\Sigma ^{(-)}$ represent sets of
initial and final states of the system, respectively. Configurations at a
surface $\Sigma ^{(+)}$ are unstable against small perturbations, and it
seems unlikely that a system could be prepared there. On the other hand, if
one considers the system at $\Sigma ^{(-)}$, a small perturbation to move it
away from the surface would require an infinite acceleration. In this sense,
the surfaces $\Sigma ^{(-)}$ represent stable final states for the evolution
of the system, and any initial configuration sufficiently near the
degeneracy surface is doomed to fall on it. Then, the question whether the
system can be consistently defined on $\Sigma ^{(-)}$ naturally arises.

For simplicity, let us consider a system possessing a single surface of
degeneracy which is globally of type $\Sigma ^{(-)}$. We will now show that
when the system reaches $\Sigma ^{(-)}$, two coordinates become non
dynamical; the system acquires a new gauge symmetry on the degeneracy
surface which corresponds to displacements along $\Sigma ^{(-)}$, and one
degree of freedom is lost.

Following Dirac's approach for constrained systems \cite{Dirac}, the action (%
\ref{action'}) possess $2n$ primary constraints coming from the definition
of the canonical momenta $p_{i}=\partial L/\partial \dot{z}^{i}$,
\begin{equation}
\phi _{i}(z,p)\equiv p_{i}-A_{i}(z,t)\approx 0\;,  \label{constraints}
\end{equation}
whose Poisson brackets are $\{\phi _{i},\phi _{j}\}=F_{ij}$. Outside $\Sigma
^{(-)}$, the invertibility of $F_{ij}$ implies that the constraints $\phi
_{i}$ are second class. However, at the degeneracy surface, the rank of $%
F_{ij}$ is reduced by two, thus, two of the $\phi $'s have vanishing Poisson
brackets with the whole set of constraints.

Although the constraint structure changes abruptly at $\Sigma ^{(-)}$, after
the system reaches this surface, its evolution can be described by a
standard constrained system, as can be seen through a suitable change of
basis for the constraints $\phi _{i}$.

Linear combinations of the form $\varphi _{(\alpha )}=e_{(\alpha )}^{i}\phi
_{i}$, become first class provided $e_{(\alpha )}^{i}$ are null vectors of $%
F_{ij}$. This can only happen at the degeneracy surface, where there are two
of such vectors. They can be chosen so that one is tangent and the other is
normal to the surfaces $F=$\ constant, namely, $e_{(1)}^{i}F_{ij}=\frac{1}{2}%
\partial _{j}F$ and $e_{(2)}^{i}F_{ij}=F_{ij}\partial _{i}\sqrt{F}$. In
Darboux-like coordinates, their only nonvanishing components are $%
e_{(1)}^{\alpha }=\epsilon ^{\alpha \beta }\partial _{\beta }f$ and $%
e_{(2)}^{\alpha }=\delta ^{\alpha \beta }\partial _{\beta }f$, with $\alpha
=1,2$.

In the basis $\phi _{i}=\{\varphi _{(\alpha )};\phi _{a}\}$, with $%
a=3,...,2n $, the constraint algebra reads,

\begin{eqnarray}
\{\varphi _{(\alpha )},\varphi _{(\beta )}\} &\approx &\frac{1}{4}\epsilon
_{(\alpha )(\beta )}F^{-\frac{1}{2}}(\partial _{i}F)^{2}=f\epsilon _{(\alpha
)(\beta )}|\partial f|^{2}\;,  \nonumber \\
\{\varphi _{(\alpha )},\phi _{b}\} &\approx &e_{(\alpha )}^{i}F_{ib}=0\;,
\nonumber \\
\{\phi _{a},\phi _{b}\} &=&F_{ab}\;.  \label{Algebra}
\end{eqnarray}
>From this it is apparent that, on the surface $\Sigma ^{(-)}$, the
constraints $\varphi _{(\alpha )}$ have vanishing Poisson brackets, and are
therefore candidates for first class constraints.

In order to examine whether $\varphi _{(\alpha )}$ are first or second class
at the degeneracy surface ($f=0$), it is necessary to compute their Poisson
brackets with $f$. The only non vanishing bracket involving $f$ is
\begin{equation}
\{f,\varphi _{(2)}\}=e_{(2)}^{\alpha }\partial _{\alpha }f=|\partial
_{\alpha }f|^{2}\;,  \label{2ndf}
\end{equation}
which cannot vanish on $\Sigma $ because, by hypothesis, $f$ has a simple
zeros there. This shows that $\varphi _{(1)}$ is first class, while, ($%
f,\varphi _{(2)}$) form a conjugate pair of second class constraints.

The transformations generated by $\varphi _{(\alpha )}$ correspond to $%
\delta z^{a}=0$, and
\begin{equation}
\delta x^{\alpha }=\{x^{\alpha },\xi ^{(\beta )}\varphi _{(\beta )}\}=\xi
^{(\beta )}e_{(\beta )}^{\alpha }=\xi ^{\alpha }\;.  \label{Translations}
\end{equation}
Thus, the constraints $\varphi _{(1)}$ and $\varphi _{(2)}$ generate tangent
and normal displacements to $\Sigma ^{(-)}$ respectively, as expected.
Hence, $f\approx 0$ can be viewed as the gauge fixing condition associated
with the ``gauge generator'' $\varphi _{(2)}$. This is summarized in the
following\newline

{\bf Lemma 4:} On the degeneracy surface $\Sigma ^{(-)}$, the system
acquires a new gauge invariance, because the second class constraint $%
\varphi _{(1)}$ becomes first class, while the number of second class
constraints ($f,\varphi _{(2)},\phi _{a}$) remains the same ($2n$). Since
each first class constraint eliminates one degree of freedom, we conclude
that one degree of freedom is dynamically frozen on the degeneracy surface.%
\newline

We illustrate these results in the following examples.

\section{Examples}

\subsection{Simplest Degenerate System}

The simplest case of a degenerate dynamical system is provided by the
Lagrangian
\begin{equation}
L_{D}=A_{\alpha }\dot{x}^{\alpha }+A_{0}\text{ },  \label{Ld}
\end{equation}
with $A_{1}=0,\;A_{2}=x_{1}x_{2},\;A_{0}=-\nu x_{1}$. The symplectic form, $%
F_{\alpha \beta }=\epsilon _{\alpha \beta }x_{2}$, degenerates at the
surface $x_{2}=0$ , which is of type $\Sigma ^{(\eta )}$, with $\eta
=sgn(\nu )$. The orbits run perpendicular to $\Sigma ^{(\eta )}$ and take a
finite time to connect a point on the surface with a point outside.

This example captures the essence of the behavior of any degenerate system
in a neighborhood of a degeneracy surface of type $\Sigma ^{(+)}$ or $\Sigma
^{(-)}$. In particular, the shock-wave solutions of Burgers' equation,
\begin{equation}
\partial _{t}u+u\partial _{x}u=\nu \partial _{x}^{2}u\;,  \label{Burgers}
\end{equation}
which is relevant in the context of turbulence, exhibit this behavior. These
solutions are of the form
\begin{equation}
u(x,t)=-2\nu \sum_{k=1}^{2n}(x-z_{k}(t))^{-1}\;,  \label{Vortex}
\end{equation}
where $z_{k}(t)\ $are complex coordinates which come in conjugate pairs and
satisfy a vortex-like equation\cite{Choodnovsky}. The corresponding
equations of motion for $z_{k}(t)$\ can be obtained from an action of the
form (\ref{action'}), which for $n=1$\ and $z=x_{1}+ix_{2}$\ reads
\begin{equation}
\left(
\begin{array}{ll}
0 & x_{2} \\
-x_{2} & 0
\end{array}
\right) \left(
\begin{array}{l}
\dot{x}_{1} \\
\dot{x}_{2}
\end{array}
\right) =\left(
\begin{array}{l}
\nu \\
0
\end{array}
\right) \;,  \label{Vortexn=1}
\end{equation}
whose associated Lagrangian, is precisely given by (\ref{Ld}). This solution
describes a one dimensional shock wave centered at $x=x_{1}$, with peaks at $%
x=x_{1}\pm x_{2}(t)$\ of height $\mp 2\nu /x_{2}(t)$, travelling outwards
from $x_{1}$.

\subsection{Coupling with a regular system}

The next example examines explicitly the fate of a degenerate system when it
reaches a surface of type $\Sigma ^{(-)}$. A simple Lagrangian for which
this occurs is of the form
\begin{equation}
L=L_{D}(x^{\alpha })+L_{R}(z^{a})-V_{\lambda }(x^{\alpha },z^{a})\;.
\label{Lt}
\end{equation}
Here,
\begin{equation}
L_{D}(x^{\alpha })=A_{\alpha }\dot{x}^{\alpha }-H_{D}(x^{\alpha })\;,
\label{Ld2}
\end{equation}
with $\alpha =1,2$, is some two-dimensional degenerate system possessing a
global surface of type $\Sigma ^{(-)}$ at $f(x^{\alpha })=0$; $L_{R}(z^{a})$
is a Regular system with Hamiltonian $H_{R}(z^{a})$, and $V_{\lambda
}(x^{\alpha },z^{a})$ is an interaction term of the form
\begin{equation}
V_{\lambda }=\lambda f(x^{\alpha })H_{R}(z^{a})\;.  \label{Coupling}
\end{equation}
This coupling is chosen so that it vanishes on $\Sigma ^{(-)}$ and does not change the flux
density $\Phi $ there, so that the character of the degeneracy surface does not depend on the
coupling constant $\lambda $. Note that this coupling would be trivial in case of nondegenerate
systems. Furthermore, the presence of $H_{R}$ in the coupling implies that, besides the
conservation of the total Hamiltonian $H=H_{D}+H_{R}+V_{\lambda }$, the equations of motion
\begin{equation}
\dot{z}^{a}=(1+\lambda f(x))F^{ab}\partial _{b}H_{R}\;,  \label{Er}
\end{equation}
give rise to a separate conservation law for $H_{R}$, because $\dot{H}_{R}=%
\dot{z}^{a}\partial _{a}H_{R}=0$. In turn, this implies that the remaining
equations of motion
\begin{equation}
\epsilon _{\alpha \beta }f(x)\dot{x}^{\beta }=\partial _{\alpha
}(H_{D}+\lambda f(x)H_{R})\text{ },  \label{Ed}
\end{equation}
can be integrated as an autonomous two-dimensional subsystem. Once these equations have been
solved, and their solutions substituted in (\ref{Er}), it is apparent that, the solutions of Eqs.
(\ref{Er}) describe the same orbits as in the decoupled case ($\lambda =0$) but with a
reparametrized time,
\[
z^{a}(t)=z_{(\lambda =0)}^{a}(\tau )\;,
\]
with
\[
\frac{d\tau }{dt}=1+\lambda f(x(t))\;.
\]
Note that as the orbits approach the surface $\Sigma ^{(-)}$, this time
reparametrization remains finite.

Once the system reaches the degeneracy surface ($f(x)\rightarrow 0$), both
time coordinates become identical and, {\em on }$\Sigma ^{(-)}${\em , all
traces of the degenerate subsystem disappear, including the information
about its initial conditions }$x^{\alpha }(t_{0})$.

Thus from the moment the degeneracy surface is reached, the system becomes a
regular one, described by $L_{R}(z^{a})$, and the degrees of freedom of the
degenerate system are forever lost.

In order to illustrate this point, consider the degenerate Lagrangian given
by Eq. (\ref{Ld}) with $\nu <0$, coupled with a one dimensional harmonic
oscillator in the form (\ref{Coupling}). In that case, the total energy is $%
{\cal E}={\cal E}_{R}(1+\lambda x_{2})+\nu x_{1}$, where ${\cal E}_{R}$ is
the energy of the harmonic oscillator, which is separately conserved. Eq. (%
\ref{Ed}) is readily integrated as
\[
x_{2}(t)=\pm \sqrt{2\nu t+(x_{2}(t_{0}))^{2}}\;,
\]
for$\;t<\frac{(x_{2}(t_{0}))^{2}}{2\nu }$, and $x_{2}(t)=0\;$afterwards.

Hence, the harmonic oscillator coordinates $Z=z^{1}+iz^{2}$ evolve according
to
\[
Z(t)=Z_{0}\exp (i\tau )\;,
\]
with $|Z_{0}|^{2}=2{\cal E}_{R}$, where the reparametrized time is given by
\[
\tau =t+\frac{\lambda }{3\nu }[2\nu t+(x_{2}(t_{0}))^{2}]^{3/2}\;,
\]
for\ $t<\frac{(x_{2}(t_{0}))^{2}}{2\nu }$, and $\tau =t$ afterwards.

\section{Discussion \& Overview}

The degeneracy of the symplectic form opens up the possibility of a violation of Liouville's
theorem. In fact, the divergence of the current $j^{i}=\sqrt{F}\dot{z}^{i}$ reads
\[
\partial _{i}j^{i}=-\partial _{i}[\sqrt{F}F^{ij}]\partial _{j}A_{0}-\sqrt{F}%
F^{ij}\partial _{i}\partial _{j}A_{0}.
\]
If $A_{0}=-H$ is continuous and differentiable, the second term in the r.h.s. vanishes
identically. However, the first term can give rise to a non-zero contribution, responsible for
the jump in the flow accross $\Sigma$. In this sense, the problem we address here is the
counterpart of Poincar\'{e} classical study of singularities in the phase flow. Both cases
correspond to different classes of possible singularities in the phase flow, and hence, the
degeneracy surfaces cannot be understood as a dense set of Poincar\'{e}'s singularities.

It is reasonable to expect that the extension of our analysis to field
theory would lead to the possibility that the symplectic form degenerates
for field configurations where some local degrees of freedom should freeze
out and some field components become nondynamical. In the case of higher
dimensional gravity, this means that as the system reaches a degeneracy
surface, some dynamical components of the metric become redundant, which
would correspond to a sort of dynamical dimensional reduction mechanism.

The quantum mechanical analysis of this kind of degenerate systems, shows that there is no
tunneling across a surface of degeneracy $\Sigma $, but there is a nonvanishing propagation
amplitude between states in the bulk and on $\Sigma $ \cite{QuantumDegenerate}. These results
would be relevant for the quantum Hall effect \cite{Asorey}, and also for strings propagating in a
background possessing a nonconstant $B$-field \cite{Seiberg-Witten}.

\noindent {\large {\bf Acknowledgments}}

We are grateful to M. Asorey, R.Bam\'{o}n, J. Cari\~{n}ena, D. Boyer, A. Gomberoff, J. Kiwi, R.
Rebolledo and C. Teitelboim for many enlightening discussions. In the initial stages of this
work, we benefited particularly from insights of I. Pi\~{n}eyro. This work was supported in part
through grants 1990189, 1010450 and 2000027 from FONDECYT, and the institutional support of a
group of Chilean companies (CODELCO, Dimacofi, Empresas CMPC, MASISA S.A. and Telef\'{o}nica del
Sur) is also acknowledged. One of us (J.S.) wishes to thank Departamento de F\'{\i}sica
Te\'{o}rica, Universidad de Zaragoza for its kind hospitality. CECS is a Millennium Science
Institute.



\begin{references}
\bibitem{Dense}  The special case when there is a dense set of
Poincar\'{e} singularities lying on $\Sigma $, is excluded by our hypotheses. In that case the
system can show a different behavior, which will be discussed elsewhere.
\end{references}
\end{document}